\documentclass[%
 reprint,
 superscriptaddress,
 frontmatterverbose, 
 amsmath,
 amssymb,
 aps,
 pre,
]{revtex4-1}

\usepackage{graphicx}
\usepackage{dcolumn}
\usepackage{bm}

\usepackage{color}
\usepackage{soul}
\begin{document}


\title{Toward an Effective Theory of Neurodynamics: Topological Supersymmetry Breaking, Network Coarse-Graining, and Instanton Interaction}
\author{Igor~V.~Ovchinnikov}
\email{igor.vlad.ovchinnikov@gmail.edu; igor.ovchinnikov@thermofisher.com}
\affiliation{R\&D, Thermo Fisher Scientific, South San Francisco}
\author{Skirmantas Janu\v{s}onis}
\email{janusonis@ucsb.edu}
\affiliation{Department of Psychological and Brain Sciences, University of California, Santa Barbara}
\begin{abstract}
    Experimental research has shown that the brain's fast electrochemical dynamics, or neurodynamics (ND), is strongly stochastic, chaotic, and instanton (neuroavalanche)-dominated. It is also partly scale-invariant which has been loosely associated with critical phenomena. It has been recently demonstrated that the supersymmetric theory of stochastics (STS) offers a theoretical framework that can explain all of the above ND features. In the STS, all stochastic models possess a topological supersymmetry (TS), and the "criticality" of ND and similar stochastic processes is associated with noise-induced, spontaneous breakdown of this TS (due to instanton condensation near the border with ordinary chaos in which TS is broken by non-integrability). Here, we propose a new approach that may be useful for the construction of low-energy effective theories of ND. Its centerpiece is a coarse-graining procedure of neural networks based on simplicial complexes and the concept of the "enveloping lattice." It represents a neural network as a continuous, high-dimensional base space whose rich topology reflects that of the original network. The reduced one-instanton state space is determined by the de Rham cohomology classes of this base space, and the effective ND dynamics can be recognized as interactions of the instantons in the spirit of the Segal-Atiyah formalism.
\end{abstract}



\maketitle


\section{Introduction}
Stochastic differential equations (SDEs) have universal applicability in science and serve as mathematical models for objects ranging in size from molecules to large-scale structures of the Universe. SDEs can also provide unique insights into the brain, one of the most complex and important known objects. In this paper, we focus on neurodynamics (ND), the basis of brain function. For the purpose of this presentation, we define ND as the neurophysiological processes associated with light-ion currents, such as spiking dynamics in neural networks.   

One important aspect of ND is that it belongs to the class of systems that exhibit the so-called self-organized criticality (SOC) \cite{Jensen_1,Zapperi_1,Bak1,PhysRevLett.78.4793,A_recent_review_on_SOC,PruessnerBook,Frigg2003613,PhysRevLett.97.118102}. SOC is a phenomenological concept that seeks to explain why a wide range of stochastic dynamical systems produce instantonic processes that obey power-law statistics, \emph{e.g.}, earthquakes and neuroavalanches \cite{Beggs03122003,Beggs02062004,ChialvoLoh}. Recently, the Parisi and Sourlas approach to Langevin SDEs \cite{Parisi_Sourlas,Parisi_Sourlas_0,ZINNJUSTIN} has been generalized to SDEs of arbitrary form \cite{Entropy}, which has allowed to recognize that the theoretical mechanism behind SOC is the spontaneous breakdown of topological supersymmetry (TS) caused by condensation of instantonic processes. 

TS is present in all SDEs. To be exact, TS is a property not of the SDE itself but rather of its stochastic evolution operator (SEO). The SEO of any SDE is of a very special form unique to the cohomological or Witten-type topological field theories (TFTs) \cite{TFTReview,Labastida_1989,Witten1,Witten2,Blau,Baulieu_1988,Baulieu_1989}. Namely, the SEO is $d$-exact, \emph{i.e.}, it equals $[\hat d, ...]$, where $\hat d$ is the exterior derivative and/or the de Rham operator that can be also recognized as the TS operator.\footnote{In high-energy physics models, TS is a more general concept.} This SEO property guarantees that all eigenstates are either \emph{supersymmetric singlets} or \emph{non-supersymmetric doublets} related to each other by $\hat d$. Furthermore, if the ground state of the model is a non-supersymmetric doublet, the TS is said to be broken spontaneously. This situation is a stochastic generalization of deterministic chaos \cite{Entropy,ChaosOrOrder,Kang}.

In SDEs, TS preserves the topology or the "proximity of points" in the phase space, \emph{i.e.}, two close initial points (or conditions) produce "close" trajectories. In deterministic systems, TS breaking allows a simple interpretation, in which close trajectories may diverge dramatically in time (this divergence can be quantified with the Lyapunov exponent). Such systems exhibit the "butterfly effect" (\emph{i.e.}, a minor change in the initial conditions may produce major differences in the outcome) and are said to be chaotic. However, this interpretation becomes untenable in stochastic systems. Specifically, the trajectory-based picture of chaos breaks down because in stochastic models \emph{all} trajectories are possible, just like in quantum theory. The proposed conceptual approach to spontaneous TS breaking and the associated Goldstone theorem can therefore generalize deterministic chaos. One important observation in this interpretation is that, in contrast to the common understanding, \emph{dynamical chaos is actually a low-symmetry state} (in the field-theoretic terminology it could be recognized as "ordered").

\begin{figure}
    \centering
    \includegraphics[width=8cm]{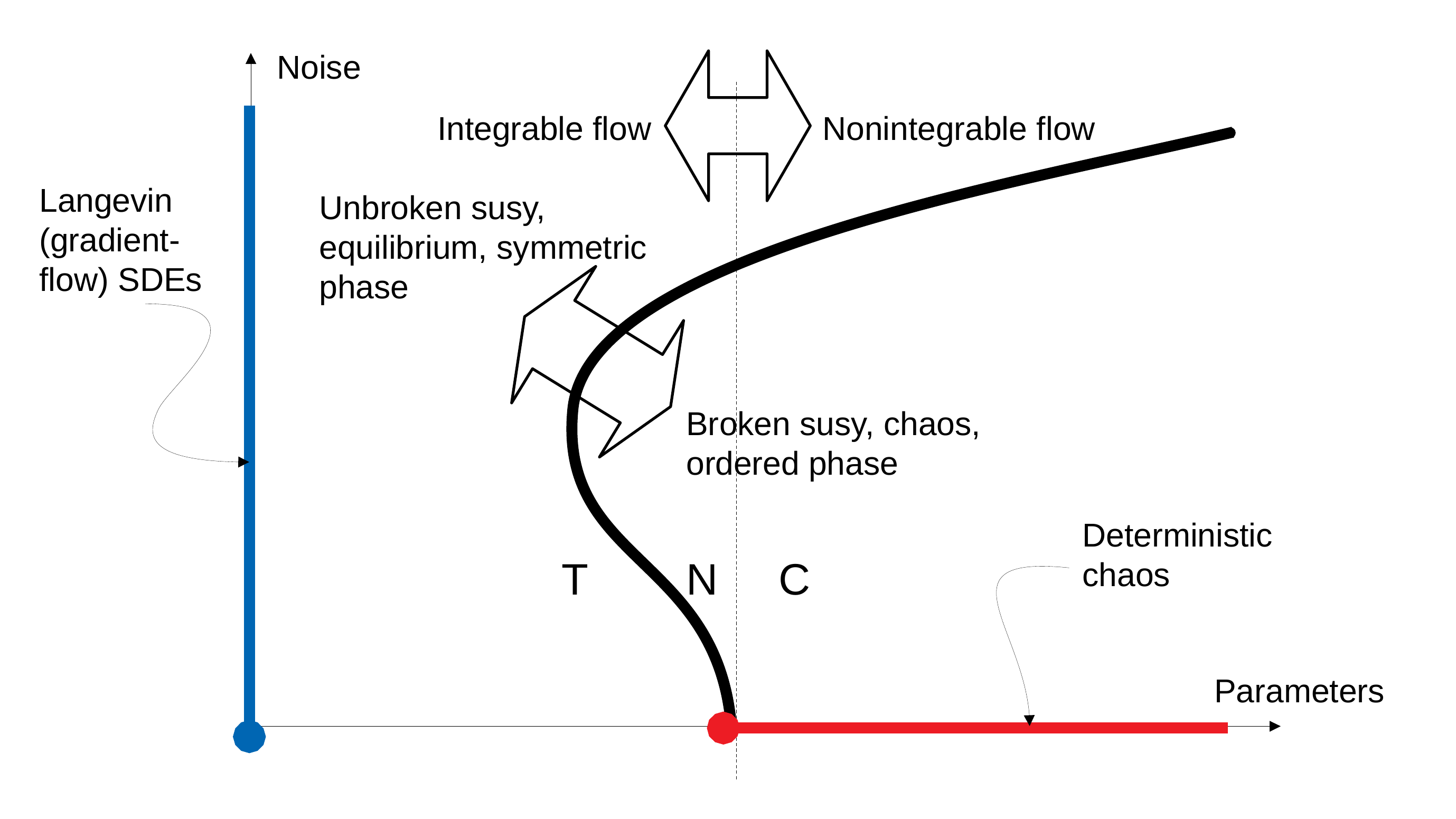}
    \caption{  The general classification of stochastic models according to the supersymmetric theory of stochastics (STS). The thick black curve separates the models with unbroken (symmetric phase, thermal equilibrium) and spontaneously broken (ordered phase, chaotic dynamics) topological supersymmetry (TS). The vertical straight line separates the models with the integrable and non-integrable deterministic part of their law of temporal evolution. Accordingly, there are three phases in the weak noise limit: the phase of ordinary chaos ($C$-phase) where TS is broken by non-integrability; the phase of noise-induced chaos ($N$-phase) where TS is broken by the condensation of instantons and noise-induced antiinstantons; and the phase of thermal equilibrium ($T$-phase) with unbroken TS. In the deterministic limit, the $N$-phase disappears. A healthy brain may reside in the $N$-phase which is covered by the STS but is beyond the scope of traditional approaches that focus exclusively on deterministic chaos (red horizontal line). The Langevin SDEs with gradient flows (blue vertical line) are studied almost exclusively in the context of the relation of stochastic dynamics to supersymmetry, but in these models TS is never broken sponetaneously. While the Martin-Siggia-Rose approach to SDEs is applicable to the entire phase diagram, it does not differentiate among the different phases.}
    \label{figure_1}
\end{figure}

In deterministic models, TS-breaking is equivalent to the concept of non-integrability (in the dynamical-system sense), which is another hallmark of the classical deterministic chaos. However, a different type of chaos may emerge in the presence of noise. In this chaos, TS can be broken by the condensation of antiinstanton-instanton configurations. This type of chaotic dynamics naturally incorporates all four key features of SOC: \emph{(i)} the dynamics is dominated by instantons (such as neuroavalanches in ND); \emph{(ii)} the self-similarity or power-law statistics of instantons can be attributed to the Goldstone theorem\footnote{The full argument that relates the power-law statistics of instantons in the $N$-phase to TS breaking can be described as follows. According to the Goldstone theorem, under the conditions of spontaneous TS breaking the low-lying fermions are gapless, which is the reason why they are sometimes called \emph{goldstinos}. Consequently, in the long-wave-length limit, goldstinos should have power-law correlators/propagators. This suggests that some characteristics of many-instanton configurations should also exhibit power-laws because, in the $N$-phase, the low-laying fermions are the supersymmetric partners of instanton/soliton moduli.}; \emph{(iii)} the position of SOC near the border of "ordinary chaos" (the $C$-phase, see Figure \ref{figure_1}) is due to noise, so that the noise-induced TS-breaking occurs before one reaches the non-integrability region associated with the $C$-phase; \emph{(iv)} in phase diagrams, the SOC regime has a finite width and therefore cannot be explained by the theory of critical phenomena (\emph{i.e.}, it is not a transition between phases but instead is a phase on its own). This makes SOC an inaccurate descriptor from the theoretical point of view; from now on, we will refer to this phase as the $N$-phase, where the "$N$" stands for "noise-induced".

As shown in Figure \ref{figure_1}, the $N$-phase is indeed unique. In particular, it is both "chaotic," because of the spontaneous breakdown of TS, and "integrable," because the deterministic part of its law of temporal evolution has not yet lost its integrablity (which would happen in the $C$-phase). Based on this uniqueness, it has been proposed that the $N$-phase dynamics is a natural optimizer \cite{SOC_optimizer}. This suggests that the $N$-phase may be important for ND. It has been proposed, based on clinical data and computer simulations \cite{Symmetry,MIT}, that the three major low-noise phases ($T$-, $N$-, and $C$-phases; Fig. \ref{figure_1}) correspond, respectively, to the coma-like, conscious-like, and seizure-like regimes of ND.

Generally, under the conditions of broken TS, unimportant fluctuations can be separated from the essential part of the dynamics. The latter can then be described by the so-called low-energy effective (field) theory (LEET). In the context of ND, an STS-based LEET would yield insights into the fundamental principles of brain function. Since the dynamics in the $N$-phase is instanton/antiinstanton-dominated, it is clear that one must focus on these objects to capture the essential properties of the system.\footnote{Note that in the $C$-phase this would not be possible because these systems are not integrable and have no distinguishable instantons.} 

 Instantons have been reasonably well studied, particularly in the context of TS and its breaking in high-energy-physics models. The neuroscience community can harness this knowledge in the construction of the ND-LEET. This paper seeks to facilitate this convergence. We are particularly interested in the \emph{base space} of coarse-grained models of ND (\emph{i.e.}, models that approximate the brain matter as a continuous medium). This base space can be assumed to be the conventional 3D-space but such a space may lose all information about the topology of the network. We take an alternative approach, in which the coarse-grained version of the base space of a neural network can be very high-dimensional and allow rich topologies. 

The structure of the paper is as follows.  In Sec. \ref{LEET_ND}, we discuss a coarse-graining procedure that deals with the topology of the neural network. We show that the de Rham cohomology classes of the base space can define the reduced one-instanton state-space of the ND-LEET, and that the overall dynamics can be understood as instaton interaction in the spirit of the Segal-Atiyah formalism. We conclude in Sec.\ref{Conclusion}. In addition, Appendix Sec.\ref{STS_general} presents a brief introduction to the key ingredients of the STS and discusses why it is advantageous over other potential approaches to stochastic dynamics. 

\section{From the STS to a Low-energy Effective Theory of Neurodynamics}
\label{LEET_ND}
In this section, we discuss the potential application of the STS to ND by using a simple model. An in-depth discussion of the STS can be found in Ref.\cite{Entropy}; for convenience, Appendix \ref{STS_general} briefly reviews the key ingredients of the STS and introduces the notation used in this section.

\subsection{Neurodynamics vs. Neuromorphics}
\label{NDvsNM}
Let us begin the discussion by first clarifying which part of brain dynamics we are focusing on. For simplification, we assume that the relatively slow dynamics (\emph{neuromorphics}, NM), associated with long-term memory and plasticity processes (including protein synthesis and transport), can be treated separately from the relatively fast dynamics (ND) of membrane electrochemical potentials, associated with small-ion flows (Na, K, Cl, Ca). The latter include such well-established neurodynamical phenomena as neuron firing, bursting, and neuroavalanches, which underlie perception and short-term memory.

This separation has a direct analogue in Artificial Intelligence where the training of neural networks can be thought of as NM and the consequent recognition (\emph{e.g.}, of new images) can be thought of as ND. Furthermore, the notable scale separation between ND and NM\footnote{ND acts on the sub-second scale, whereas NM can operate over hours, days, and years.} makes it reasonable to use an "adiabatic" approximation, where ND can be studied at a fixed NM configuration.\footnote{Note that this allows Bayesian perception. The real situation is of course much more complex. For example, brain oscillations have a nested structure \cite{Raichle_2010} and can span frequencies from $>500\text{Hz}$ to slow oscillations (with a period of around 1 min). Also, waking consciousness is strongly supported by slow processes, such as the circadian rhythm.} While not perfect, this separation of time-scales allows a divide-and-conquer strategy. 

\subsection{The reduced state space for N-phase dynamics}

If TS is spontaneously broken in a system, the meaningful part of its dynamics (\emph{i.e.}, the one that does not include unstructured fluctuations) occupies a reduced state/phase space. Take magnetism for instance. There, the reduced state space of a system that contains a huge number of electrons is the space of all configurations of only one field, the magnetization (order parameter). In this context, the corresponding LEET is known as micromagnetodynamics, described by the Landau-Lifshits-Gilbert equation. A similar approach may be possible in ND. Again, the first step on the way to a ND-LEET requires the identification of a reduced state space.

To set the stage, let us recall that ND of the healthy brain is likely to reside in the $N$-phase. The established characteristics of ND are well reflected in the theoretical properties of this phase. Specifically, for an external observer the dynamics looks like a sequence of instantons (neuroavalanches), and some features of these instantons obey power-laws (see, \emph{e.g.}, Refs.\cite{ThresholdProcesses,10.3389/fnsys.2015.00022,Plenz_2019} and Refs. therein).

From a more technical point of view, instantons are condensed (into the ground state) and cause spontaneous breakdown of the TS in the $N$-phase. Instantons, in turn, represent \emph{transitions} between different perturbative, or \emph{local}, supersymmetric ground states associated with \emph{local unstable manifolds} of the flow $\mathcal F$ (see the Appendix for a brief discussion of local unstable manifolds).\footnote{In the $N$-phase, the flow is still integrable and local unstable manifolds are still well-defined topologically. In the deterministic limit, the vacua are the Poincar\'e duals of the local unstable manifolds. This contrasts with the $C$-phase, where local unstable manifolds acquire a fractal-like structure and can fold on themselves in a recursive manner.} \footnote{In models with a nontrivial base space, instantons are processes of destruction of solitons such as kinks, vortices, domain-walls, \emph{etc.}. Accordingly, antiinstantons are processes of creation of solitons by noise. Each vacuum represents an unstable solitonic configuration. Therefore, the state in which the instantons are condensed into the ground state, \emph{i.e.}, the $N$-phase, can be pictured as a state with noise-induced dynamics and soliton interactions, \emph{i.e.}, boundaries of neuroavalanches in ND.} They are also called \emph{vacua} in high-energy physics, and we adopt this nomenclature here.

In the approximation that neglects inter-vacua transitions (\emph{i.e.}, instantons/antiinstantons), all vacua are supersymmetric and have zero-eigenvalues. These vacua also have "towers of states" above them that represent fluctuations. Therefore, if only the vacua are considered, these fluctuations are essentially disregarded. If the goal is an effective description of the instanton/antiinstanton dynamics, this approach is reasonable and the vacua can be viewed as a reduced state space of the ND-LEET.

When the instantons/antiinstantons are taken into account, the reduced SEO (in the basis of the vacua) acquires non-zero off-diagonal matrix elements (instantonic matrix elements). The reduced SEO is related to the scattering matrix discussed in Sec.\ref{Atiyah_Segal}. In the $N$-phase, the diagonalization of the reduced SEO provides a \emph{global} ground state which is a superposition of multiple vacua and is non-supersymmetric (\emph{i.e.}, its TS will be spontaneously broken by the instantons).

\subsection{A prototypical firing event in zero-dimensions}
\label{1_neuron_section}
\begin{figure*}
    \centering
    \includegraphics[width=15cm]{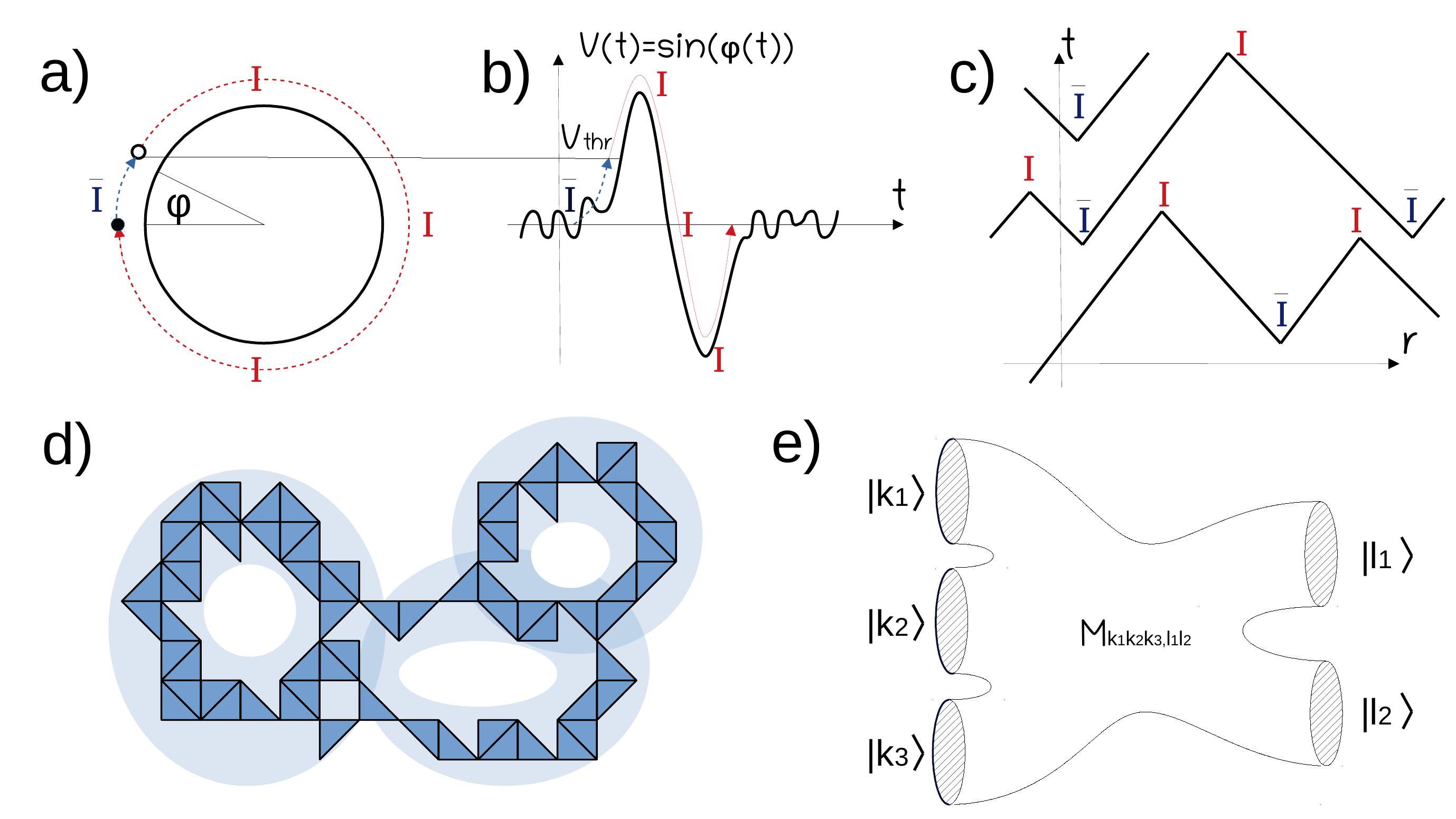}
    \caption{{\bf (a)} The phase space of the single-neuron model described in Eq.(\ref{0D_SDE}) is $S^1$. The flow has a stable fixed point (attractor) at $\varphi=0$ (small filled circle) and an unstable fixed point close to it (small empty circle). The firing event consists of an antiinstanton ($\bar I$) and an instanton ($I$). The antiinstanton is produced by noise and moves the system against the flow vector field ($\mathcal F$) from the attractor to the unstable fixed point (and thus over the threshold (thr)). The instanton returns the system back to the attractor along the flow. {\bf(b)} The $\bar I$-$I$ pair in terms of the "membrane voltage" ($V=\sin(\varphi)$). Again, the firing event consists of the $\bar I$ excitation above the threshold (dashed line at $V_{thr}$) and the subsequent $I$ that returns the neuron to its resting potential. {\bf (c)} In the 1D-extension of the model described by Eq.(\ref{1Dmodel}), the base space (without time) is a large spatial circle. In this case, the antiinstanton is the noise-induced creation of a kink-antikink pair (solitons at which the field jumps by $\pm2\pi$). Once created, the kink and antikink propagate in the opposite directions. When kinks and antikinks from different antiinstantons meet, they annihilate each other in an instantonic process. The black lines denote the positions of kinks and anti-kinks as functions of time. {\bf (d)} A simple example of the coarse-graining procedure of a 2D-neural network, as discussed in the text. The graph is first embedded into a sufficiently high-dimensional (square) lattice in such a way that only nearest neighbors are connected by synapses. The base space (shaded area) is the interior of the co-dimension $1$ surface tightly wrapped around the simplicial complex (collection of triangles) constructed from the graph. For realistic neural networks, the base space would be ultra-high dimensional and have a very rich topology. {\bf (e)} Instanton interaction can be approximately understood using the Atiyah-Segal formulation of topological/conformal field theory, \emph{i.e.}, as a functor from the category of cobordisms to the category of linear spaces. The borders of the surface that represent the scattering matrix are the model's base space (without time). The linear spaces associated with each border are one-instanton vacua tagged by the de Rham cohomology classes of the base space. There are in- (right hatched circles) and out- (left hatched circles) one-instanton states.}
    \label{figure_2}
\end{figure*}

Let us introduce instantons using a simple model (Fig.\ref{figure_2}a,b) that can be thought of as a simplified description of an isolated neuron \cite{Symmetry}. The model has only one dynamical variable, $\varphi \in S^1$, whose evolution in time is described by 
\begin{subequations}
\begin{eqnarray}
\frac{d \varphi}{d t} (t) = {\mathcal F}(t|\varphi(t)) = \alpha - \left.\frac{\partial U_{0D,sG}}{\partial \varphi}\right|_{\varphi = \varphi(t)} + (2\Theta)^{1/2}\eta(t),
\end{eqnarray}
\label{0D_SDE}
where $U$ is the zero-dimensional sine-Gordon potential     
\begin{eqnarray}
U_{0D,sG}(\varphi) = -\cos(\varphi + \varphi_0),
\end{eqnarray}
so that 
\begin{eqnarray}
\frac{\partial U_{0D,sG}}{\partial \varphi}(\varphi) = \sin(\varphi + \varphi_0).
\end{eqnarray}
\end{subequations}

The variable $\varphi$ can be thought to represent the membrane potential, with the parameter $\alpha$ as its threshold parameter. The constant $\varphi_0$ is introduced for convenience and set at $\varphi_0 = \sin^{-1}\alpha$, so that at $\varphi=0$ the neuron is at its resting state. This point is a stable critical point (or an attractor) of the deterministic flow. In addition, the flow has an unstable critical point at $\varphi_1 = \pi - 2\varphi_0$. The unstable critical point is the tip of the barrier that the neuron has to overcome to fire. The values of $\alpha$ that are relevant to ND are those where $1-\alpha = \delta\ll 1, \varphi_0 = \pi/2-\epsilon, \epsilon\ll 1$. For such values, the saddle is close to the attractor ($\varphi_1-\varphi_0=2\epsilon\ll1$). (We note that when $\alpha>1$, the model loses both of its fixed points. In higher-dimensional versions of the model, $\alpha=1$ corresponds to the loss of integrability and the onset of chaotic/turbulent dynamics even in the deterministic case.)

The above model represents as a class of models that link Langevin/potential dynamics with the sine-Gordon potential in the limit $\alpha\to0$ and with Kuramoto oscillators\footnote{For the use of Kuramoto oscillators in the context of ND see, \emph{e.g.}, Ref.\cite{Kuramoto_2020} and Refs. therein.} in the limit $\alpha\to\infty$.  

If noise is weak, the dynamics looks as follows. Most of the time, $\varphi$ fluctuates around its attractor. Once in a while, noise pushes the system over the barrier; $\varphi$ then undergoes a full-circle rotation around its phase space ($S^1$) and comes back to the attractor from the other side. This firing event is a predecessor of neuroavalanches. The full circle trajectory along $S^1$ has two parts: the excitation from the resting state to the saddle and the return from the saddle to the resting state along the long arm of $S^1$. 

These two parts are antiinstanton and instanton processes, respectively.\footnote{One straightforward analogue of the firing event is popping a baloon with a needle. There, piercing the surface of the balloon is the antiinstanton whereas the collapse that produces the popping sound is the instanton.} The antiinstanton is the noise-driven dynamics \emph{against} the flow vector field (the corresponding matrix element has exponentially weak Gibbs-like factors that vanish in the deteministic limit). The instanton, on the other hand, is the motion \emph{along} the flow vector field that needs no assistance from noise. 

As discussed in the previous section, the two critical points host two vacua which are the Poincar\'e duals of the unstable manifolds of the two critical points. They are respectively the delta-functional distribution,
\begin{eqnarray}
|0\rangle = \delta(\varphi-\varphi_0)\chi,
\end{eqnarray} 
where $\delta$ is the Dirac delta-function, and the constant function, $C=1$, over the entire phase space,
\begin{eqnarray}
|1\rangle = C. \label{saddle_vacuum}
\end{eqnarray}
The extended character of the second vacuum reflects the presence of a modulus in the class of solutions that lead from the saddle to the resting state. In this case, the modulus is the time of occurrence of the instanton. In higher-dimensional versions of the model, instanton moduli are higher dimensional and they also include the position in the base space where the firing event was initiated. 

\subsection{Neuroavalanches as instantons}
\label{Neuroavalanche_Instanton}
Before we proceed, it is important to briefly review why a neuroavalanche can be recognized as an instanton-associated process. First, it must be pointed out that in a more general setting relevant to Morse-Smale dynamical systems, instantons can be defined as families of deterministic solutions of a static flow that connect critical points of different stability in the spirit of Morse theory. In this very general concept, any classical trajectory with this distinct feature can be recognized to an extent as an instanton. However, neuroavalanches can be linked to a narrower understanding of instantons, as processes that destruct solitons in the setting of nonlinear sigma (or spatially extended) models. 

Let us recall that a neuroavalanche is a chain-reaction of neuron firing where one firing neuron causes firing in neurons that are connected to it, with further propagation. If neurons were theoretically located on, say, a cubic lattice with only nearest neighbors connected, a neuroavalanche would look like a growing “bubble” where the surface of the moving “domain wall” would separate the pre- and post-firing neurons. 

Neural networks are of course not cubic lattices and the actual propagation of neuroavalanches is much more complex. If, however, we could “straighten out” the network by embedding it into a higher-dimensional “enveloping” lattice with only nearest neighbor-coupling (see Sec.2.6 below for details), the neuroavalanche would look like a moving domain wall in this lattice. 

Once neuroavalanche is recognized as a moving domain wall (a soliton), it only remains to notice that in extended dissipative systems (such as nonlinear sigma models) processes of destruction of solitons are called instantons. This destruction happens by the annihilation of the soliton with an anti-soliton, as in the case discussed below in Sec. \ref{1D_sineG}. Alternatively, in finite extended systems (such as neural networks) a soliton can disappear by reaching the spatial boundary of the system. Either way, solitons move to a position where they get destructed, and this motion is a part of the instantonic process.

We conclude that a neuroavalanche can be identified as an instantonic process of the propagation and destruction of the neuronal soliton that separates pre- and post-firing neurons in the higher-dimensional “enveloping” space. The process of the creation of the neuronal soliton can be called a neuronal antiinstanton, so that the entire process becomes the creation, propagation, and destruction of a single isolated neural soliton can be treated as an antiinstanton/instanton pair. We discuss these processes further below.


\subsection{An overdamped sine-Gordon model in one dimension}
\label{1D_sineG}

The one-neuron model in one of the previous subsections \ref{1_neuron_section} has a finite-dimensional phase space and a trivial (zero-dimensional) base space (see Fig.\ref{figure_2}c). The simplest generalization of this model is a 1D-chain of neuron-like elements which are coupled to their nearest neighbors. This model can be coarse-grained into a spatially extended model with a continuous 1D-base space such as a circle. Accordingly, the 1D-generalization of Eq. (\ref{SPDE}) has the form 
\begin{subequations}
\label{1Dmodel}
\begin{eqnarray}
\frac{\partial \varphi}{\partial t} (rt) &=& {\mathcal F}(rt|\varphi(t)) \nonumber \\&=& \alpha - \left.\frac{\delta U_{1D,sG}}{\delta \varphi(r)}\right|_{\varphi = \varphi(t)} + (2\Theta)^{1/2}\eta(rt),
\end{eqnarray}
where $r$ is the spatial coordinate of the base space and the one-dimensional sine-Gordon potential is     
\begin{eqnarray}
U_{1D,sG}(\varphi) = \int dr \left((\partial_r\varphi(r) )^2/2 - \cos(\varphi(r) + \varphi_0)\right),
\end{eqnarray}
so that
\begin{eqnarray}
\frac{\delta U_{1D,sG}}{\delta \varphi(r)} = -\partial_r^2 \varphi(r) + \sin(\varphi(r) + \varphi_0).
\end{eqnarray}
\end{subequations}

As in the previous section, the parameter $\varphi_0 = \sin^{-1}\alpha$ is introduced for convenience so that at $\varphi(rt)=0$ all neurons are at their (stable) rest state. 

Equation (\ref{1Dmodel}) can be viewed as a non-potential ($\alpha>0$) extension of the overdamped sine-Gordon equation \cite{sine_Gordon_Kink_Bath, Josephson_general} or the Frenkel-Kontorova equation \cite{Review_Frenkel_Kontorova}. Such models are used, for instance, for 1D-chains of Josephson junctions \cite{PhysRevB.54.1234, Josephson_general} where they describe the temporal evolution of voltages. As mentioned in the previous section, the non-potential part $\alpha$ relates Eq. (\ref{1Dmodel}) to Kuramoto oscillators which are used in the context of ND \cite{Kuramoto_2020}.\footnote{The Kuramoto limit ($\alpha\to\infty$) in oscillating neurons, \emph{i.e.}, neurons that fire non-stop, should corresponds to the seizure- or $C$-phase of ND, as discussed in the Introduction.}

The elementary firing event, or the antiinstanton/instanton pair, is the creation and consequent annihilation of a pair of \emph{solitons} called kink and antikink (for review of solitions see, \emph{e.g.}, Refs.\cite{Soliton_classics, Solitions_Review_1, sine_Gordon_Kink_Bath} and Refs. therein). These solitons are the spatially localized rotation of $\varphi$ by $\pm 2\pi$. In the \emph{antiinstanton} part of the firing event, the \emph{kink-antikink pair} is created by noise out of the stable vacuum. The kink and antikink start moving away from each other at a constant speed (related to the constant $\alpha$). When the solitons meet on the opposite side of the spatial circle, they annihilate each other in the \emph{instanton} part of the firing event.

It is important to point out that the analogue of the unstable critical point in the previous section is a saddle point in the configuration where the kink and antikink are balanced (position-wise) in the sense that the driving force pushing them in the opposite directions is compensated for by the attraction force between them. We note that this balance condition only fixes the \emph{relative} positions of the solitons and that there are actually an infinite number of such configurations (the mutual positions of balanced solitons is arbitrary). This spatial position of the kink-antikink pair is yet another instanton modulus which represents the point at which the kink-antikink pair was created.

With regard to the flow vector field, what was a saddle critical \emph{point} in the 0D-model in the previous section now becomes a saddle critical \emph{manifold} (circle), which is just the base space of the model. This brings the discussion into the domain of Morse-Bott theory \cite{Review_Bott_Morse_Theory}. The latter says that the vacua associated with critical manifolds belong to the de Rham cohomology of these manifolds. Instead of a single vacuum in Eq. (\ref{saddle_vacuum}), now there are two vacua from the zeroth and first cohomology of the base space:
\begin{eqnarray}
|k\rangle = C(\sigma) \wedge \theta_{k}(R) \wedge f(\triangle\varphi_{\perp}), \theta_{k} \in H_k, k=0,1. \label{k_1D_sinG}
\end{eqnarray}
Here, the function $C$ is the direct analogue of the same function in Eq.(\ref{saddle_vacuum}), $\sigma$ is the time-like modulus, $\theta$ is the Morse-Bott factor from the two cohomology classes of the base space, $R$ is the coordinate or spatial modulus, and $f$ is yet another factor which is the very narrow distribution in all of the other transverse modes that can be defined as
\begin{eqnarray}
\varphi(tr) \approx \varphi_{I}(rt|\sigma R) + \triangle\varphi_\perp(rt|\sigma R),
\end{eqnarray}
where $\varphi_{I}(rt|\sigma,R)$ is the "classical" one-instanton configuration and $\triangle\varphi_\perp$ is the transverse fluctuations around the instanton/unstable manifold,
\begin{eqnarray}
\int drdt \triangle\varphi_{\perp}(rt)\cdot \frac{\partial \varphi_{I}(rt|\sigma R)}{\partial z} = 0, z=\sigma,R. \end{eqnarray}

\subsection{The coarse-grained base space of ND}
\label{coarsegraining}

The above model can be straightforwardly generalized to models with more complicated base spaces. Therefore, we are now ready to discuss the base space that can be used for the LEET of ND.

Naturally, neural networks can be viewed as embedded in the conventional physical 3D-space. However, they also can be viewed as extremely large and complex graphs\footnote{Neural networks are also directed graphs because action potentials travel in one direction and chemical synapses are fundamentally asymmetric. An ND-LEET with a continuous base space may allow mimicking the directional character of the graph (\emph{e.g.}, by the introduction of a gauge field that makes some directions preferred for the propagation of solitons/avalanches). This topic is outside the scope of this paper.}\cite{Book_Networks_Sporn}. This property may allow coarse-graining. In particular, the dynamically important distance between two neurons in the network can be assumed to be the least number of links that connect them, instead of their separation in the 3D-anatomical space (if the dynamics of the propagation of electrochemical signals in cellular compartments of individual neurons is neglected\footnote{Neurons vary dramatically in their morphologies, axon lengths, and other spatial characteristics which can affect temporal properties of networks. Interestingly, neurons may functionally 'normalize' some of these morphological variations \cite{Hamada_2016,Schmidt_2019,Parajuli_2020,Verbist_2020}. }).

The coarse-graining procedure we propose reflects both the topology and geometry of the network (Fig. \ref{figure_2}d). In order to find an acceptable coarse-graining procedure, let us recall that in solid state physics a crystal lattice at larger scales is coarse-grained into a \emph{continuous} 3D-space because at shorter distances the lattice looks the same at every point and only nearest neighbors are linked through the hopping matrix elements. This hints at the first step of the required coarse-graining which may embed the network into a sufficiently high-dimensional lattice. Let us call it the \emph{enveloping lattice}. The dimensionality of this lattice, $D_L$, must be high enough to make sure that the synapses connect \emph{only the nearest neighbors} on the lattice. It is clear that the number of nearest neighbors on the lattice ($2D_L$ for a square lattice) must be larger than the largest number of synaptic connections produced by a single neuron. Therefore, the practical values of $D_L$ are on the order of tens of thousands.

The enveloping lattice can correctly treat the \emph{geometry} of the network (the interneuronal distances in the graph). In order to correctly reflect the \emph{topology} of the network, we can think of the network as a \emph{simplicial complex} (for a review, see, \emph{e.g.}, Ref.\cite{Nakahara}). This basic topological concept is useful for many purposes. In our case, it helps to correctly fill the "interior" of the graph. The rule for including a simplex into the graph is the following: if $N$ neighboring neurons are all connected to one another (\emph{i.e.}, form a clique), they can be considered an $N$-simplex and included, with all of the simplex's lower-dimensional edges, into the simplicial complex.

Finally, the base space can now be thought of as a volume enclosed by a $D_L-1$ surface, tightly wrapped around the simplicial complex. This procedure leads to a very high dimensional, "spongy" base space that has a very rich topology (which represents the original network). For the simple case of $D_L=3$, this procedure is schematically shown in Figure \ref{figure_2}d.

\subsection{Instanton interaction and the Atiyah-Segal formulation}
\label{Atiyah_Segal}

In section \ref{1D_sineG}, we discussed one-instanton vacua. These states can be said to define the internal states of isolated instantons or, using the field-theoretic nomenclature, their scattering states that do not interact in the infinite future/past. Just like in Eq.(\ref{k_1D_sinG}), let us denote these states as
\begin{eqnarray}
|k\rangle.
\end{eqnarray}
This time, however, $k$ runs over the de Rham cohomolgy classes of the coarse-grained base space that represents the neural network (as discussed in Sec.\ref{coarsegraining}). We note that the number of the indices is very large ($1\le k\le N, N \gg 1$).

It follows from the discussion in Sec.\ref{1D_sineG} that the number of Paddeev-Popov ghosts of a one-instanton state is
\begin{eqnarray}
f(k) = d(k)+1,\label{numb_fermion}
\end{eqnarray} 
where $d(k)$ is the degree of the cohomology class, $k$, and the unity on the right side of the equation comes from the fundamental instanton modulus that can be associated with the time of the occurrence of the instanton.

These states are the building blocks for the overall dynamics in the $N$-phase that can be described as interacting instantons. Many-instanton interactions, in turn, can be defined by the following scattering matrix:
\begin{eqnarray}
M(k_1...k_m|l_1...l_n) = \langle k_1...k_m |  {\mathcal M}_{+\infty,-\infty} |l_1...l_n\rangle,
\end{eqnarray}
where $\mathcal M$ is the stochastic evolution operator defined in Eq.(\ref{SEO}) and
\begin{eqnarray}
|l_1...l_n\rangle = |l_1\rangle \otimes ...  \otimes |l_n\rangle,
\end{eqnarray}
where we use the bra-ket notation for the $n$ one-instanton vacua.

The number of fermions must be conserved in scattering processes. This leads to the conclusion that the scattering amplitudes are described by
\begin{eqnarray}
&M(k_1...k_m|l_1...l_n) \sim m(k_1...k_m|l_1...l_n)) \times \nonumber \\ &\times \delta_{\sum_{i=1}^m f(k_i), \sum_{j=1}^n f(l_j)},
\end{eqnarray}
where $\delta$ is the Kronecker delta and the functions $f$ are defined in Eq.(\ref{numb_fermion}).

The concept of the scattering matrix is intuitively clear because it is built on the traditional field-theoretic picture of interacting particles. Nevertheless, its rigorous definition is not straightforward. For example, the very idea that instantons do not interact in the infinite future/past is clearly an idealization that can be reliable only in some well-defined situations. A careful treatment of this problem falls outside the scope of this presentation. Instead, in order to provide the scattering matrix a more formal look we can turn to the Segal-Atiyah formalism. It views the scattering matrix as a functor from the category of \emph{cobordisms} into the category of \emph{vector spaces}, as illustrated in Fig.\ref{figure_2}e.

The Segal-Atiyah formalism has been proposed for the conformal and topological field theories where the vector Hilbert spaces are finite-dimensional. The same is true in our case because the number of cohomology classes is finite (here, we are interested not in the entire theory but only in its application to the approximation of instanton dynamics). As any approximation, it has applicability limits and is mostly appropriate for the $N$-phase dynamics with dilute instantons (which is most likely realizable in the low-noise limit).

It should be noted that the scattering matrix essentially plays the role of \emph{a new stochastic evolution operator}. The diagonalization of this operator can reveal the low-lying spectrum of the model and, in particular, can answer the question of whether the TS is broken or not. Further work in this direction will depend crucially on the approximations used for the construction of the scattering amplitudes, which falls outside the scope of this discussion.

\section{Conclusion}
\label{Conclusion}

In this paper, we discuss the key elements of the ND-LEET, from the perspective of the supersymmetric theory of stochastic dynamics. The noise-induced chaotic regime, which is believed to host the ND of the normal brain, is characterized by the spontaneous breakdown of TS by configurations of instantons and noise-induced antiinstantons. Accordingly, the LEET must describe gapless Faddeev-Popov ghosts which are the supersymmetric partners for the moduli (global collective variables) of these configurations. At the same time, the ghosts must be directly related to the topology of the neural network.

We propose the following setting that satisfies the above requirements. The centerpiece of our proposition is a coarse-graining procedure of the neural network that turns it into a \emph{high-dimensional continuous base space} with a topology that reflects the original network configuration. The base space is an intrinsic part of the one-instanton modulus and, in the spirit of Morse-Bott theory, the de Rham cohomology classes represent the "quantum number" of the corresponding one-instanton states/vacua. Accordingly, the reduced base space of the ND-LEET is a multi-instanton extension of the one-instanton vacua. Consequently, the dynamics of neuronal avalanches can be described in terms of interacting instantons, and the corresponding scattering amplitudes can be conveniently understood using the Atiyah-Segal formalism. 

We outline only the first step. A number of questions will have to be answered before the STS picture of ND can become practically useful. These include but are not limited to the calculation of the (anti-)instanton matrix elements between the vacua, the diagonalization of the effective Hamiltonian, the determination of the order parameter, the analysis of the response of the system to external stimuli/perturbations, and the understanding of how these components are linked to the (already existing) information-processing concepts. However, this effort may be well justified. Specifically, \emph{information processing and storage in ND may well have a topological character} which would explain its robustness to internal and external perturbations. 

We also note that the proposed approach can contribute to fundamental neuroscience more broadly. Noisiness is often thought to be deleterious to biological systems (including the brain) which should overcome it in order to achieve robust self-organization and homeostatic states. However, early studies have shown that the self-organization of molecular gradients, an essential step in neurodevelopment, can be easily achieved in simple reaction-diffusion systems, \emph{provided the initial molecular concentrations are noisy} \cite{Babloyantz_1977}. More recently, fluctuations in molecular biosystems have been demonstrated to be important for efficient control \cite{Hilfinger_2016}. Also, the brain contains many axons that produce fundamentally stochastic trajectories. These axons, classically placed in the vaguely defined "ascending reticular activating system," have now been shown to allow rigorous descriptions based on anomalous diffusion processes \cite{Janusonis_2020,Vojta_2020}. Intriguingly, these "stochastic axons" routinely interact with the classical "deterministic axons" (which connect specific brain structures and are studied in the connectomics framework). They are well-positioned to support the $N$-phase (Fig. \ref{figure_1}), thus extending the proposed conceptual approach into NM, beyond fast ND processes. Therefore, the discussed framework may contribute to many areas of neuroscience.            

\acknowledgments{This research was supported by the National Science Foundation (grant \#1822517 to SJ), the National Institute of Mental Health (grant \#MH117488 to SJ), and the California NanoSystems Institute (Challenge grants to SJ). }




\appendix

\section{The Supersymmetric Theory of Stochastics}
\label{STS_general}

\subsection{The class of models}
We consider a class of models that covers the model discussed later in the paper, as well as other potential ND models:
\begin{eqnarray}
\partial_t \varphi(rt)  = {\mathcal F}(rt| \varphi(t)) + (2\Theta)^{1/2} \eta (rt),\label{SPDE}
\end{eqnarray}
where $t\in \mathbb{R}^1$ is time, $r$ is the spatial coordinate from the base space $\mathcal B$ (which we assume to be a topological manifold \footnote{Our use of the term 'base space' is different from that in high-energy physics models where base spaces also include time} ), $\varphi$ is the dynamical variable/field from the target space ($\mathcal T$), $\eta$ is Gaussian white noise uniquely defined by its fundamental correlator $\langle \eta(x)\eta(x')\rangle = \delta(x-x')$, $\Theta$ is the intensity of the noise, and $\mathcal F$ is the flow vector field that represents the deterministic part of law of evolution and which is a functional of the field configuration ($\varphi(t)$). This functional may explicitly depend on $r$ and $t$ (here and below, we use "$rt$" instead of "$r, t$" for brevity; we also introduce a combined space-time coordinate, $x=(r, t))$.

In the traditional approach to stochastic dynamics, the temporal evolution of the system is described by the Fokker-Planck evolution of the probability distribution over the phase space. In the class of models under consideration, the phase space, $X$, is the infinite-dimensional space of all (sufficiently smooth) configurations of the field over the base space, or, equivalently, all the maps from the base space to the target space (\emph{i.e.}, $X = \{ \varphi | \varphi:{\mathcal B}\to{\mathcal T}\}$). Accordingly, the probability distribution is a function(al) on $X$.

\subsection{The Markovianity and reduced information in the probability distribution}
One important aspect of Focker-Planck evolution is its Markovianity: the instantaneous change of the probability distribution function(al), $P(t)$, depends only on $P(t)$ at this time\footnote{This is true for models with white noise. We consider only such models.}:
\begin{eqnarray}
\partial_t P(t|\varphi) = - \hat H P(t|\varphi),
\end{eqnarray}
where $\hat H$ is the Fokker-Planck operator that will be defined later.

There are infinitely many pasts of the system, including perturbations, that lead to a predetermined $P(t)$. Therefore, $P(t)$ contains only reduced information about the system's past. Specifically, the classical descriptions in terms of $P$ are limited by the information capacity of $P$, as there may exist systems in which $P$ is not "sufficiently large" to contain all pertinent information about the system's history. Chaotic systems are quintessential examples of such systems in that they may have an infinitely long memory of the initial conditions and perturbations during their evolution. An extension of the theory from $P$ to some more general "wavefunctions" is needed to treat such systems correctly. 


The use of generalized wavefunctions may require some justification and it is in this context that it is relevant to discuss briefly the meaning of the classical probability distribution. The probabilistic description is a natural choice for a stochastic model because neither the initial conditions nor the configuration of the noise term are known exactly to the observer. At the same time, it is clear that in reality a given physical dynamical system has concrete initial conditions and experiences only one specific realization of noise. Therefore, it has a concrete position in its phase space at each time-point.

The resolution of this apparent contradiction is straightforward. The probability distribution does not describe the physical dynamical system directly. Instead, it describes the external observer or rather his knowledge about the physical system. It can therefore be said that unlike, say, the wavefunction in quantum mechanics which is often considered to be a physical object \cite{PhysicalWavefunction}, $P$ is not a physical but rather a mathematical object, a representation of the external observer's knowledge about the system and its past. From this point of view, switching to other mathematical coordinate-free objects is not a big leap, especially since these objects have a clear probabilistic meaning within the context of conditional probability densities.

\subsection{A generalization to differential forms and the potential physicality of STS ghosts}

The traditional probability distribution is not the only entity that can represent an abstract external observer. For example, an observer may have conditional information about the system. This kind of information can be represented as conditional probabilities which, in turn, can be given as differential forms. For our class of models, this would mean that instead of the probability distribution we could consider a "wavefunction,"
\begin{eqnarray}
P(t|\varphi) \to \psi(t,\varphi\chi),
\end{eqnarray}
where $\chi(r)$ is a fermionic field or Faddeev-Popov ghosts (see below) represented by functional differentials over the phase space, $\chi(r)\sim\wedge \delta \varphi(r)$.

One major advantage of this class of wavefunctions is that their geometro-topological character allows establishing the law of their temporal evolution with relative ease. Namely, the SEO that propagates the wavefunctions forward in time can be defined as the result of the stochastic averaging of the action (also known as pullback) induced by the SDE-defined maps (the collection of all trajectories). As discussed, for example, in detail in Ref.\cite{Entropy}, the SEO is given by
\begin{widetext}
\begin{eqnarray}
{\mathcal M}_{t_ft_i}(\varphi_{f}\chi_{f}|\varphi_{i}\chi_{i} ) = \iint_{o.b.c.}{\mathcal D} \Phi \; e^{S_{t_ft_i}(\Phi)} = e^{-(t_f-t_i)\hat H} \delta(\varphi_f-\varphi_i)\delta(\chi_f-\chi_i),\label{SEO}
\end{eqnarray}
\end{widetext}
where the functional integration is over $\Phi(x)=(\varphi(x),B(x),\chi(x),\bar\chi(x))$, which is the collection of the original field, $\varphi$, the Lagrange multiplier or momentum, $B$, and the pair of Faddeev-Popov ghosts, $\chi,\bar\chi$\footnote{This field can be recognized as Faddeev-Popov ghosts because stochastic quantization can be interpreted as a gauge fixing procedure \cite{TFTReview}. Accordingly, the operator in (\ref{TS}) can also be recognized as the operator of Becchi-Rouet-Stora-Tyutin supersymmetry of this gauge-fixing procedure.} (with $x=(r,t)$). The functional integration is over all paths with open boundary conditions (o.b.c.) that connect the initial and final configurations of the supersymmetric extension of the field, $\varphi(x)|_{t=t_f,t_i} = \varphi_{f,i}(r)$, $\chi(x)|_{t=t_f,t_i} = \chi_{f,i}(r)$. 

The action in Eq.(\ref{SEO}) is given by
\begin{eqnarray}
S_{t_ft_i}(\Phi) = \{ {\mathcal Q}, \Psi_{t_ft_i}(\Phi)  \},\label{action_Q_exact}
\end{eqnarray}
where the TS operator is given by 
\begin{eqnarray}
{\mathcal Q} = \int dx \left(\chi(x) \frac{\delta}{\delta \varphi(x)} + B(x) \frac{\delta}{\delta\bar \chi(x)}\right),\label{TS}
\end{eqnarray}
where $\int dx = \int_{t_i}^{t_f} dt \int_{\mathcal B} dr$ and the so-called gauge fermion is given by
\begin{eqnarray}
\Psi_{t_ft_i}(\Phi) = \int dx \; i\bar\chi(x)\left(\partial_\tau \varphi(x) - {\mathcal F}(x| \varphi(t)) \right. \nonumber \\\left.+ \Theta i B(x)\right).
\end{eqnarray}
The Fokker-Planck operator in Eq.(\ref{SEO}) is given by
\begin{eqnarray}
\hat H = [\hat d, \hat {\bar d} ], \label{SEO_d_exact}
\end{eqnarray}
where the exterior derivative and "current operator" are
\begin{eqnarray}
\hat d &=& \int_{\mathcal B} dr \chi(r)\frac{\delta }{\delta \varphi(r)}, \\ 
\hat{\bar d} &=& \int_{\mathcal B} dr \frac{\delta}{\delta\chi(r)}\left({\mathcal F}(r|\varphi) - \Theta \frac{\delta }{\delta \varphi(r)}\right). 
\end{eqnarray}
The ${\mathcal Q}$-exact action in Eq.(\ref{action_Q_exact}) and the $\hat d$-exact SEO in Eq.(\ref{SEO_d_exact}) are a unique feature of the cohomological or Witten-type topological field theories.

Recall that we argued in the previous section that the extension of the Hilbert space from the classical probability distribution to differential forms is justified in part by the non-physicality of the probability distribution. But how "real" are the Faddeev-Popov ghosts?   

We note that the STS and its ghosts offer a dual TFT description of the SDE. This description, however, leads to a solid definition and stochastic generalization of chaos, as well as an explanation for SOC. It is reasonable to suspect that this duality is more than a mathematical trick. 

For example, in dynamical systems theory, one can numerically find the Lyapunov exponents by evolving differentials along the trajectory of the chaotic model. These differentials are "real" for specialists in dynamical systems theory and they are also exactly ghosts in the STS framework. 

When TS is broken spontaneously in chaotic systems, the ground state has a nontrivial ghost content (as will be discussed in the next section). \emph{It can be said that these ground-state ghosts are the price to pay for a Markovian description of a chaotic model that retains infinite memory of its initial conditions and perturbations.} 

In terms of the path-integral representation of evolution, the fermionic ghosts represent the functional determinant of the original bosonic fields of the SDE\footnote{Consistent with the nomenclature of particle physics, we call the original field of the SDE, $\varphi$, \emph{bosonic} to counterpose them to \emph{fermionic} ghosts. It should be pointed out that not all features of quantum particles have their counterparts in the STS. Nevertheless, the differentials are indeed equivalent to fermionic fields from the mathematical point of view \cite{Witten_1982}.}, while the chaotic memory (or "the butterfly effect") is the long-range order that this determinant develops under the conditions of spontaneous TS breaking. Thus the ghosts of the ground state can also be viewed as collective fields that represent the chaotic memory of the model.

The ghosts have been originally introduced only for mathematical convenience. However, history shows that some objects that were once believed to be purely abstract mathematical inventions have turned out to be considerably more physical or even observable. One prominent example is the Bohm-Aharonov experiment that has revealed the physicality of the vector potential of the magnetic field and of the phase of the phase function. Therefore, it is not outside the realm of possibilities that the STS ghosts may eventually turn out to be more than just a mathematical sleight-of-hand. In the meantime, we find it convenient to think of them as \emph{quasiparticles that are directly linked to the information contained in the dynamical system}. If the methods for detecting ghosts are found in the future, they are likely to be based on the recently proposed pipeline that can determine whether a signal is chaotic or not \cite{Toker}.

\subsection{Supersymmetry and its spontaneous breakdown}

In physics, symmetries lead to protected degeneracies of the eigenstates of evolution operators. In other words, the eigenstates come in multiplets which are irreducible representations of a particular symmetry group and the eigenstates from the same multiplets have exactly the same eigenvalue. In the case of TS, the symmetry divides all eigenstates into supersymmetric singlets and non-supersymmetric doublets of the form $|\zeta \rangle, \hat d|\zeta\rangle$. Each de Rham cohomology of the phase space hosts one supersymmetric eigenstate with an eigenvalue that is exactly zero. All the other eigenstates are non-supersymmetric pairs. Their eigenvalues are real or complex conjugate pairs, with the real part bounded from below.

When one of such eigenstates is the ground state, defined as the fastest growing eigenstate, the TS is said to be broken spontaneously. The deterministic counterpart of this phenomenon is known in dynamical system theory as dynamical chaos and/or as non-integrability of the flow vector field, $\mathcal F$. The real part of the eigenvalue of the non-supersymmetric ground state is related to dynamical entropy and the sum of the negative Lyapunov exponents via the Pesin entropy formula.

In deterministic models, the non-integrability of the flow is the only way how TS can be broken spontaneously. In the presence of noise, however, TS can also be broken spontaneously by condensation of instantons and noise-induced antiinstantons that match them. In less technical terms, antiinstanton/instanton configurations are noise-induced tunneling processes between, say, different attractors. 

To elaborate on this important concept, the term instantons is borrowed from high-energy physics where it denotes, in most cases, a more specific class of nonlinear objects than the one we use here. In our more general case, instantons are (the families of) solutions connecting critical points of the deterministic flow. From the physical point of view, every "transient" process such as neuroavanche, earthquake, or a bubble in a boiling water can be recognized as an instanton. In the spatialliy extended models, instantons are processes of removal of solitons by means of annihilation with antisoliton or by means of pushing solitons out of the system.

When supersymmetry is spontaneously broken in this way, it can be said that the model is the noise-induced chaotic (or $N$-phase) because in the deterministic limit antiinstantons (noise-induced processes) disappear together with this phase. On a phase diagram, the noise-induced chaotic phase "precedes" ordinary chaotic behavior. At low noise intensities, this phase can be sustained but remains narrow, which is why it has been believed for a long time that the $N$-phase has something to do with critical behavior. This, however, contradicts the finite width of this phase (Fig.\ref{figure_1}). The STS-based view resolves this issue.

The STS picture of $N$-phase dynamics is particularly important in ND because the healthy brain is likely to operate in this phase.

\section{Instantons, unstable manifolds, and perturbative supersymmetric ground states}


One convenient way to introduce vacua and instantons is through the concept of (un)stable manifolds. To simplify the discussion, let us first discuss "gradient-like" flows with isolated critical points of $F$, $\{x_\alpha| F(x_\alpha)=0\}$ that contain no invariant cycles. The stable/unstable manifolds, $M_{\uparrow,\downarrow}(x_0)$, associated with a critical point, $x_0$, consist of all points of the phase space that flow under $\dot x(t)=F(x(t))$ to $x_0$ at $t \to \pm \infty$. It is clear that $dim M_{\downarrow}(x_0) = codim M_{\uparrow}(x_0) = ind (x_0)$, where the index of the critical point $ind(x_0)$ is the number of its unstable directions, \emph{i.e.}, the eigenvectors of the matrix $\partial F^i/\partial x_j$ with a negative real part in their eigenvalues.

Now, the instanton connecting two critical points $x_{\alpha}$ and $x_{\beta}$ is the overlap, $I_{\alpha,\beta} = M_{\downarrow}(x_\alpha) \cap M_{\uparrow}(x_\beta)$. This formula can be interpreted as follows: the instanton between the two critical points is the family of all ODE solutions that start at one critical point and end at the other. It is clear that $dim I_{\alpha\beta} = ind (x_\alpha) - ind (x_\beta)$, which implies $ind(x_\alpha) > ind (x_\beta)$ because the solutions always move from a less stable critical point to a more stable one.

The concept of (un)stable manifolds is also useful because these manifolds are closely related to the concept of locally supersymmetric ground states. Let us introduce the Poincar\'e duals of unstable manifolds, $\bar M_{\downarrow}(x_\alpha)$, which are differential forms such that $\int \psi\wedge\bar M_{\downarrow}(x_\alpha)  = \int_{M_{\downarrow}(x_\alpha)}\psi$, where $\psi$ is any differential form of degree $ind(x_\alpha)$: $\psi\in\Omega^{(ind(x_{\alpha}))}$, where $\Omega^{(k)}$ is the space of all differential forms of degree $k$.

\section*{References}
\bibliographystyle{apsrev4-1}
\bibliography{LEET_ND.bib}



\end{document}